\documentclass[prl,twocolumn,showpacs,superscriptaddress,amsfonts,amsmath,floatfix]{revtex4}
\usepackage{graphicx}
\usepackage[dvipdfm]{hyperref} 
\usepackage{verbatim}
\usepackage[normalem]{ulem}
\usepackage{color}

\newcommand{\arcctg}{\mathop{\rm arcctg}\nolimits}
\newcommand{\arctg}{\arctan}

\newcommand{\g}{g_{1\text{D}}}
\renewcommand{\a}{a_{1\text{D}}}

\begin{document}
\title{Trapped one-dimensional ideal Fermi gas with a single impurity}
\author{G.E. Astrakharchik}
\affiliation{Departament de F\'{\i}sica i Enginyeria Nuclear, Campus Nord B4, Universitat Polit\`ecnica de Catalunya, E-08034 Barcelona, Spain}
\author{I. Brouzos}
\affiliation{Institut f\"ur Quanteninformationsverarbeitung, Universit\"at Ulm, 89069 Ulm, Germany}
\date{\today}

\begin{abstract}
Properties of a single impurity in a one-dimensional Fermi gas are investigated in homogeneous and trapped geometries.
In a homogeneous system we use McGuire's expression [J. B. McGuire, J. Math. Phys. {\bf 6},  432  (1965)] to obtain interaction and kinetic energies, as well as the local pair correlation function.
The energy of a trapped system is obtained (i) by generalizing McGuire expression
 (ii) within local density approximation
(iii) using perturbative approach in the case of a weakly interacting impurity and (iv) diffusion Monte Carlo method.
We demonstrate that a closed formula based on the exact solution of the homogeneous case provides a precise estimation for the energy of a trapped system for arbitrary coupling constant of the impurity even for a small number of fermions.
We analyze energy contributions from kinetic, interaction and potential components, as well as spatial properties such as the system size.
Finally, we calculate the frequency of the breathing mode. Our analysis is directly connected and applicable to the recent experiments in microtraps.

\end{abstract}

\maketitle

Ultracold Fermi gases in the uniform space and in harmonic traps have been studied both theoretically\cite{review_theory} and experimentally\cite{review_exp},
exhibiting fascinating phenomena such as the crossover from BCS superfluid\cite{bcs_exp} to Bose-Einstein condensate (BEC) of molecules\cite{mol_exp}, or FFLO states in imbalanced Fermi mixtures\cite{fflo}.
Some of the recent experiments have focused on the case of large population imbalance which leads to the polaronic behavior for attractive\cite{polaron_exp_att} and repulsive impurity atoms\cite{polaron_exp_rep}.
The results are found to be in a good agreement with corresponding theoretical treatments\cite{polaron_theory_att,polaron_theory_rep}.
The latter experiments performed in three-dimensional (3D) traps are commonly in a weakly interacting regime where the mean-field uniform theory is valid.
On the contrary, experiments done in quasi one-dimensional (1D) tubes\cite{kinoshita,haller} or microtraps\cite{selim1,selim2}
have been able to demonstrate unusual phenomena of strongly interacting 1D gases, like the Tonks-Girardeau (TG) limit\cite{girardeau}
where the local correlation function is totally suppressed $g_2=0$\cite{kinoshitag2}. Recently many theoretical works discuss Fermi-mixtures in lower dimensions \cite{fermi1d}.
In advanced experiments with microtraps\cite{selim1,selim2} a few-body 1D Fermi-mixture is prepared with each component consisting of a specific number of fermions.
Currently\cite{selim}, the ground-state energy of an ideal Fermi gas with a specific number of atoms interacting with an impurity was measured with a high precision for different interaction strengths.

In this Letter we provide a theoretical treatment of 1D trapped ideal Fermi-gas system interacting via a contact potential of arbitrary strength with an impurity.
We set off from the corresponding uniform case for which an analytical expression for the ground-state energy was derived by McGuire\cite{mcguire}, and we analyze the energy contributions and the pair correlation function.
We demonstrate that a modification of this solution offers a precise expression to the ground-state energy in the trap, comparing with exact numerical results
and perturbative methods.
Additionally, we perform an analysis of the energy contributions and calculate the frequency of the breathing mode, which is another important observable for the actual experiments.
Our treatment is directly applicable to the ongoing experiments \cite{selim} and provides a solid theoretical ground for understanding the properties of impurities in 1D trapped fermionic systems or similar physical models.

The system of a 1D ideal Fermi gas with an impurity
(usually an atom of the same species with equal mass $m$ in a different hyperfine state eg. for $^6$Li $|F=1/2, m_F= \pm 1/2\rangle$ \cite{selim1,selim2})
is described by a simple and quite general model Hamiltonian:
\begin{eqnarray}
\label{ham}
\hat H =
-\frac{\hbar^2}{2m} \sum_{i=1}^N \frac{\partial^2}{\partial x_i^2} + \frac{m \omega_{\parallel}^2}{2} \sum_{i=0}^N x_i^2
-\frac{\hbar^2}{2m}\frac{\partial^2}{\partial x_{imp}^2}
 \qquad \\ \nonumber
+ \frac{m \omega_{\parallel}^2}{2} x_{imp}^2 +
\g \sum_{i=1}^N \delta(x_i-x_{imp})\;,
\end{eqnarray}
where $x_{imp}$ is the position of a single impurity, $x_i$ with $i=\overline{1,N}$ are positions of ideal fermions, and $\g$ is impurity-fermion coupling constant.
In the experiments in quasi-1D traps\cite{kinoshita,haller,selim1,selim2}
the transversal trapping frequency $\omega_{\perp}$ is typically one order of magnitude larger than the longitudinal one $\omega_{\parallel}$ validating the 1D treatment
(the trap becomes then highly anisotropic with the transversal oscillator length $a_{\perp} \equiv \sqrt {\hbar/m \omega_{\perp}}$
much smaller than the longitudinal one $a_{\parallel} \equiv \sqrt {\hbar/m \omega_{\parallel}}$).
The scattering properties of the system are modified by the trapping geometry leading to confinement-induced resonances\cite{olshanii}
and are described effectively by a single parameter, the 1D $s$-wave scattering length $\a = - \left(1-\frac{|\zeta(1/2)| a_{3\text{D}}}{\sqrt{2} a_{\perp}}\right)a^2_{\perp}/a_{3\text{D}}$,
where $a_{3\text{D}}$ is the 3D s-wave scattering length (tuned by magnetic field via Feshbach resonances).
Due to antisymmetry, $s$-wave interactions between identical fermions are not possible, while $p$-wave interactions can be neglected.
Therefore the only relevant interaction in the system acts between the impurity and identical fermions and can be modeled by a zero-range $\delta$-potential with interaction strength $\g=-2\hbar^2/m\a$\cite{olshanii}.

\emph{Homogeneous system:}
In the uniform case, $\omega_{\parallel}=0$, the dynamics are governed by the competition between the kinetic and the interaction energy.
The length scales of this problem are the size of the box $L$, the mean interparticle distance $\rho^{-1} = L/N$ [or the inverse Fermi momentum $k_F^{-1} = (\pi\rho)^{-1}$] and $s$-wave scattering length $\a$.
In the thermodynamic limit $L\to\infty$ the size of the box drops out and the only relevant parameter left is the dimensionless Lieb-Liniger parameter $\gamma = -2/(\rho \a)$.

\begin{figure}
\includegraphics[width=\columnwidth, angle=0]{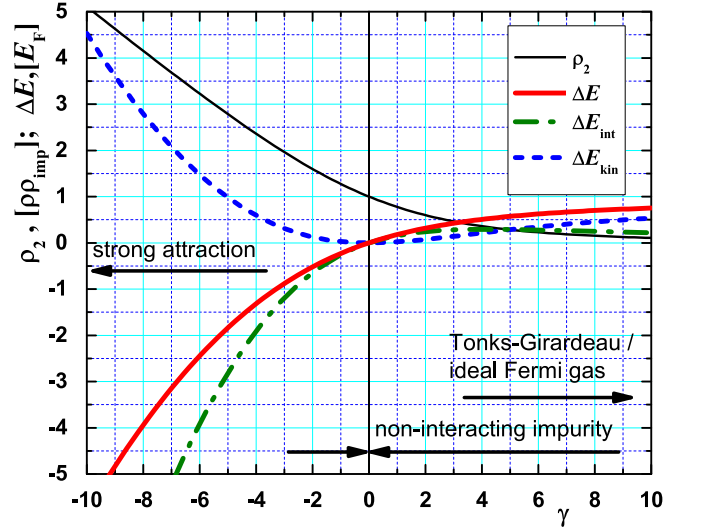}
\caption{(Color online) Thin line: impurity-fermion local pair correlation function $\rho_2$ as given by Eq.~(\ref{Eq:g2hom}).
Thick lines show the interaction shift to the energy in units of Fermi energy: red solid line, total energy $\Delta E$, Eq.~(\ref{Eq:Ehom});
green dash-dotted line, interaction energy $\Delta E_{int}$, Eq.~(\ref{E:int:hom}); blue short-dashed line, kinetic energy $\Delta E_{kin}$, Eq.~(\ref{E:kin:hom}).}
\label{Fig1}
\end{figure}

The ground-state energy $E$ of a homogeneous system was found analytically by McGuire\cite{mcguire}
in a form of an interaction shift $\Delta E$ to the energy of an ideal Fermi gas $E = N E_F/3 + \Delta E$ 
with
\begin{eqnarray}
\frac{\Delta E}{E_F}
=\frac{\gamma}{\pi^2}
\left[1 - \frac{\gamma}{4}+\left(\frac{\gamma}{2\pi}+\frac{2\pi}{\gamma}\right)\arctg\frac{\gamma}{2\pi} \right],
\label{Eq:Ehom}
\end{eqnarray}
where $E_F=\hbar^2 k_F^2/2m$ is the Fermi energy.
This expression is valid both for repulsive and attractive interactions\cite{mcguire}.
We will calculate the local pair correlation function\cite{kinoshitag2}
$\rho_2 = \langle \hat\Psi^\dagger(x)\hat\Psi_{imp}^\dagger(x)\hat\Psi_{imp}(x)\hat\Psi(x)\rangle$ (where $\hat\Psi(x)$ and $\hat\Psi_{imp}(x)$ are fermion and impurity field operators, respectively)
which is proportional to the probability of observing simultaneously the impurity and a fermion at position $x$.
In a homogeneous system $\rho_2$ does not depend on the position and its local value can be calculated applying the Hellmann-Feynman theorem\cite{Gangardt03}
$dE/d\g = \left\langle d\hat H/d\g \right\rangle = \rho_2 L$ leading to
\begin{eqnarray}
\frac{\rho_2}{\rho\rho_{imp}} = 1+\frac{\gamma}{2\pi}
\left(
\arctg\frac{\gamma}{2\pi} - \frac{\pi}{2}
\right)\;,
\label{Eq:g2hom}
\end{eqnarray}
where $\rho_{imp}=\langle\hat\Psi_{imp}^\dagger\hat\Psi_{imp}\rangle=1/L$ is the density of the impurity.
As shown in Fig.~\ref{Fig1}, the pair correlation function decreases continuously as the interaction strength $\gamma$ increases.
It vanishes in the limit of infinitely strong repulsion $\gamma\to+\infty \Rightarrow \rho_2 \to 0$, i.e., the impurity avoids meeting other fermions which mimics Pauli exclusion principle.
In the opposite limit of strong attraction the pair correlation function diverges $\gamma\to-\infty \Rightarrow \rho_2 \to +\infty$, i.e., the impurity forms a zero-size deep bound state and its position coincides with the position of one of the fermions.
In the case of vanishing interactions, $\gamma=0$, the fermions and the impurity are uncorrelated and therefore $\rho_2=\rho \rho_0$ reduces to a simple product of the corresponding densities.

The interaction energy $\Delta E_{int}$ is related to the pair correlation function~(\ref{Eq:g2hom}) as $\Delta E_{int} = \g \rho_2 L$ resulting in
\begin{eqnarray}
\frac{\Delta E_{int}}{E_F}
=\frac{2\gamma}{\pi^2}
\left[1 - \frac{\gamma}{4}+\frac{\gamma}{2\pi}\arctg\frac{\gamma}{2\pi} \right]
\label{E:int:hom}
\end{eqnarray}
The same result can be obtained from the Hellmann-Feynman theorem as $\Delta E_{int} = \g\langle d\hat H / d\g\rangle$.
On the other hand, the shift in the kinetic energy $\Delta E_{kin}$ due to the interaction between the impurity and the fermions is given by the difference between the total energy~(\ref{Eq:Ehom}) and the interaction energy~(\ref{E:int:hom}):
\begin{eqnarray}
\frac{\Delta E_{kin}}{E_F}
=\frac{\gamma}{\pi^2}
\left[-1 + \frac{\gamma}{4}+\left(\frac{2\pi}{\gamma}-\frac{\gamma}{2\pi}\right)\arctg\frac{\gamma}{2\pi} \right]
\label{E:kin:hom}
\end{eqnarray}

Figure~\ref{Fig1} shows the dependencies of different contributions to the energy shift on the interaction parameter $\gamma$.
In the case of strong attractive forces, $\gamma\to-\infty$, the large interaction energy leads to a strong bounding between the impurity and a fermion.
The interaction energy provides the main contribution at intermediate repulsive coupling $\gamma \approx 3$ but, approaching the TG limit ($\gamma\to +\infty$), it follows the decrease of the probability of contact ($\rho_2$) and the gas becomes ideal with $N+1$ fermions, i.e. the impurity is \emph{fermionized}.

\emph{Trapped system:}
In the presence of a trap, $\omega_{\parallel} > 0$, the exact energy is not known,
but we will show that an expression based on the solution of the homogeneous system [Eq.~(\ref{Eq:Ehom})] provides a very precise description.

First we treat the problem within local density approximation (LDA).
The chemical potential $\mu^t$ of a trapped system is approximated as a sum $\mu^t = \mu + \frac{1}{2}m\omega_{\parallel}^2x^2$ of the homogeneous chemical potential $\mu$ and the external potential.
The density profile of ideal fermions has a semicircular shape $\rho_{LDA}(x) = \frac{m \omega_{\parallel}}{\pi\hbar} \sqrt{R^2-x^2}$ with the size of the cloud $R$ such that $\mu^t = m\omega_{\parallel}^2 R^2/2$.
The total number of fermions in the trap defines the normalization conditions 
and fixes the chemical potential to $\mu^t = N\hbar\omega_{\parallel}$.
The chemical potential $\mu^t$ can be interpreted in terms of the Fermi momentum of a trapped system $k_F^t$ as $\mu^t = \hbar^2(k_F^t)^2/2m$ with $k_F^t = \sqrt{2N}/ a_{||}$.
It is interesting to note that the energy of $N$ ideal Fermi particles $N^2\hbar\omega_{\parallel}/2$, obtained within LDA, coincides with the exact result.
We will look for the energy shift $\Delta E= E-E_0^t$ to the energy of a non-interacting impurity  $E_0^t=\hbar \omega_{\parallel} (N^2+1)/2$.

A first analytic expression for $\Delta E$ can be obtained perturbatively in the case of weak interactions, taking for the impurity a Gaussian (non-interacting) profile
$\rho_{imp}(x_{imp}) = \frac{1}{\sqrt{\pi} a_{||}}\exp(-x_{imp}^2/a_{||}^2)$ and using the LDA. The energy shift can be calculated as $\Delta E = \int\!\int\rho_{LDA}(x)\rho_{imp}(x_{imp}) \g\delta(x-x_{imp})dxdx_{imp} = \int\rho_{LDA}(x)\rho_{imp}(x)\g dx$ which results in the expression
\begin{eqnarray}
\label{E:weak interaction:full}
\frac{\Delta E}{\hbar\omega_{||}} &=&-\frac{2 N a_{||}}{ \pi \a} [I_0(N)+I_1(N)] e^{-N}\\
&\approx& -\frac{2\sqrt{2N} a_{||} }{\pi\a} +O(N^{-1/2})
\label{E:weak interaction:expansion}
\end{eqnarray}
where the modified Bessel functions of first kind $I_0,I_1$ can be expanded for large $N$ resulting in a $\sqrt{N}$ dependence.

An explicit expression for $\Delta E$ valid for arbitrary interaction strength can be obtained by generalizing McGuire's expression for the energy~(\ref{Eq:Ehom}) of a homogeneous system to the trapped case according to the mapping $k_F \to k_F^t$,
that is the Fermi momentum of a homogeneous system is substituted by the Fermi momentum of a trapped system.
Accordingly, we introduce a characteristic parameter $\gamma^t$ of a trapped system in analogy with the Lieb-Liniger parameter $\gamma = -2\pi / (k_F\a)$.
In a trap the Fermi momentum changes to $k_F^t = \sqrt{2N}/a_{\parallel}$ and therefore we define $\gamma_t$ as $\gamma_t =  -\sqrt{2 /N}\;\pi a_{\parallel}/ \a$.
The resulting energy in terms of $\gamma_t$ is
\begin{eqnarray}
\frac{E}{\hbar \omega_{\parallel}} =
\frac{N^2\!+\!1}{2}
+ \frac{N\gamma_t}{\pi^2}
\left[1 - \frac{\gamma_t}{4}+\left(\frac{\gamma_t}{2\pi}+\frac{2\pi}{\gamma_t}\right)\arctg\frac{\gamma_t}{2\pi} \right]
\label{E:dim}
\label{Etrap:dim}
\end{eqnarray}

We test this expression by confronting it with numerically exact results of diffusion Monte Carlo (DMC) method\cite{boronat}.
This method is based on solving the Schr\"odinger equation in imaginary time and asymptotically gives the exact ground-state energy.
The convergence is greatly increased by using the importance sampling according to the guiding wave
$\psi(x_1, ..., x_N, x_{imp}) = \exp(-\alpha x_{imp}^2) \prod_{i=1}^N \exp(-\beta x_i^2) \prod_{j<k}^N |x_j-x_k| \prod_l |x_l - x_{imp} - \a|$,
where parameters $\alpha$ and $\beta$ are optimized by minimizing the variational energy.

\begin{figure}
\includegraphics[width=0.9\columnwidth, angle=0]{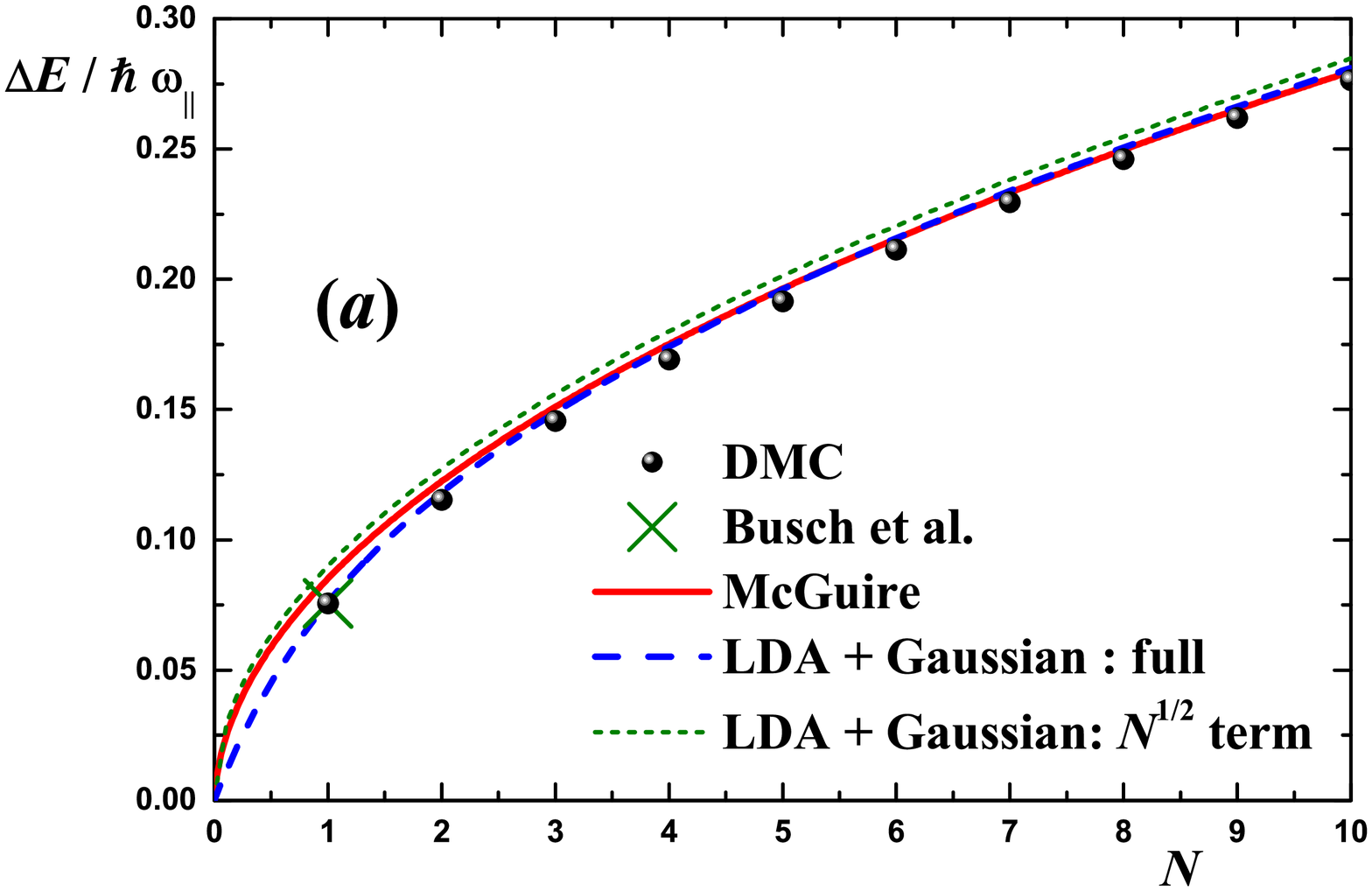}
\includegraphics[width=0.9\columnwidth, angle=0]{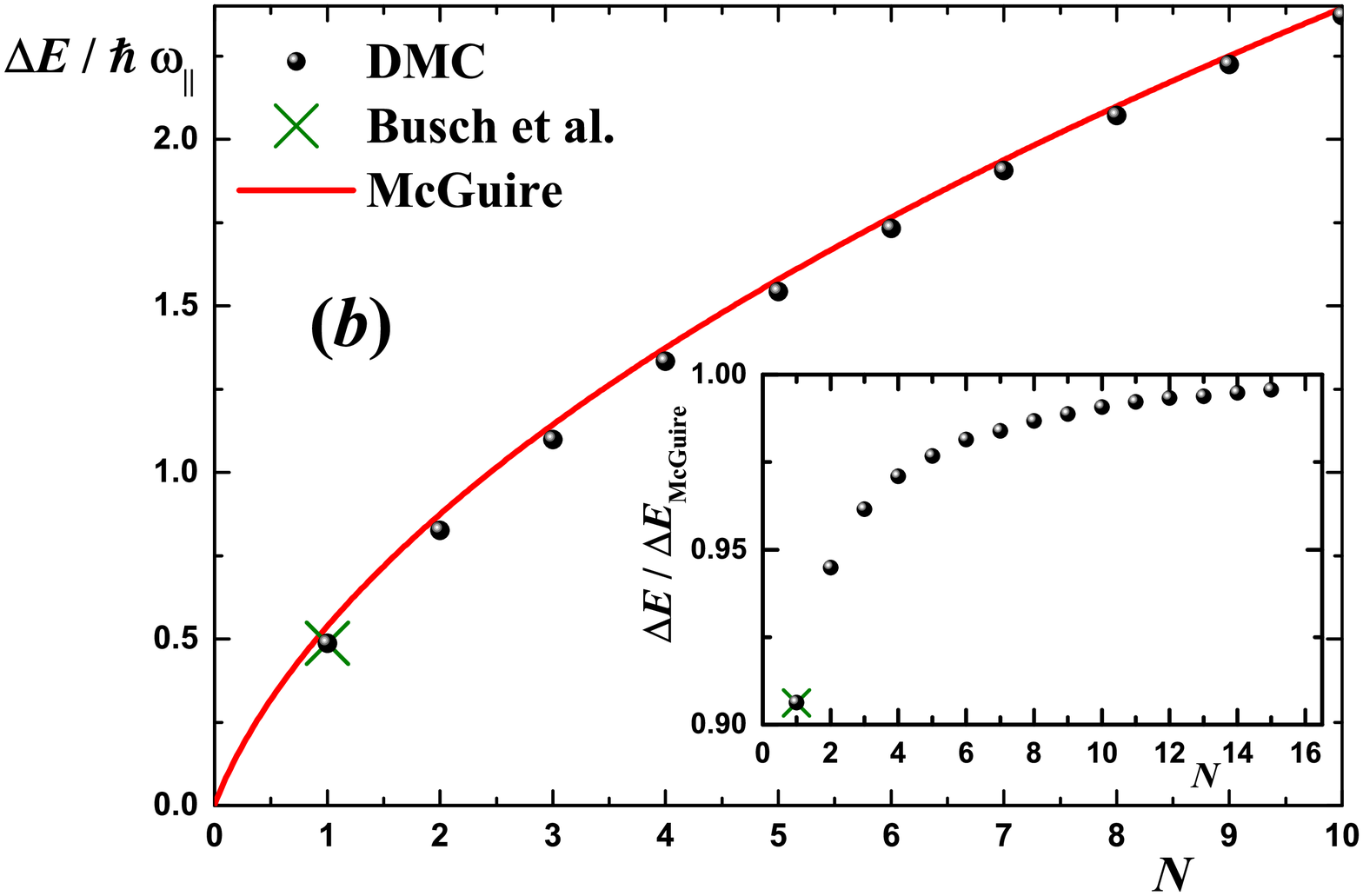}
\caption{(Color online) Ground-state energy of $N$ fermions and a single impurity in a trap in oscillator units. (a) weak $\a/a_{||}=-10$ and (b) strong $\a/a_{||}=-1$ interaction.
Circles: DMC points;
cross: exact two-particle (one fermion, one impurity) energy from Ref.~\cite{busch};
red solid line: modified McGuire expression~(\ref{Etrap:dim});
blue dashed line: perturbative expression with as Gaussian impurity + LDA , full expression, Eq.~(\ref{E:weak interaction:full});
green short-dashed line: the same, $\sqrt{N}$ expression, Eq.~(\ref{E:weak interaction:expansion}).
Inset in (b): DMC energy shift versus energy shift in modified McGuire expression~(\ref{Etrap:dim}).
}
\label{Fig2}
\end{figure}

The accuracy of the different approaches is analyzed in Fig.~\ref{Fig2} where dependence of the energy of a trapped system on $N$ for different interaction strength is confronted with the exact results of DMC calculations.
In the case of weak interactions, $\a/a_{||}=-10$ [Fig.~\ref{Fig2}a], the  modified McGuire energy~(\ref{Etrap:dim}) becomes exact for $N \gg 1$,
while for few fermions a certain difference is present as also can be seen from comparison with the exact result of Busch et al. for $N=1$\cite{busch}.
The perturbative calculation agrees well in its full form, Eq.~(\ref{E:weak interaction:full}), and slightly less accurate in $\sqrt{N}$ expansion, Eq.~(\ref{E:weak interaction:expansion}).
For stronger interaction strength, $\a/a_{||}=-1$ [Fig.~\ref{Fig2}b], the modified McGuire energy works well for large number of fermions $N$.
In order to quantify the convergence of this expression on $N$ we study the correction in the inset of Fig.~\ref{Fig2}b.

Once we have demonstrated an excellent accuracy of the modified McGuire's formula, we use it to calculate the contributions to the total energy.
The potential energy of the harmonic confinement is related to the mean square system size which is calculated using Hellmann-Feynman theorem $\frac{\partial E}{\partial \omega_{\parallel}^2} =\frac{1}{2} m \langle x^2 \rangle$:
\begin{eqnarray}
\frac{E_{pot}}{\hbar \omega_{\parallel}} =
\frac{1}{2} \frac{ \langle x^2\rangle} {a_{||}^2}=
\frac{N^2+1}{4}
+\frac{N}{\pi}
\arcctg \frac{\gamma_t}{2\pi}
\label{Epot:dim}
\end{eqnarray}
In the limit of vanishing interactions $\gamma_t\to 0 \Rightarrow \langle x^2 \rangle/a_{||}^2 = (N^2+1)/2$ the size of the system is that of a non-interacting impurity and $N$ ideal fermions.
The infinitely repulsive (TG) case $\gamma_t\to +\infty \Rightarrow \langle x^2 \rangle/a_{||}^2 = (N+1)^2/2$ is equivalent to having $N+1$ ideal fermions.
The size in the limit of an infinitely strong attraction, $\gamma_t\to -\infty \Rightarrow \langle x^2 \rangle/a_{||}^2 = (N-1)^2/2$, is that of $N-1$ fermions since one fermion is absorbed in the bound state with the impurity,
which as well can be seen from the separation of the energy~(\ref{E:dim}) into binding and ideal Fermi gas energies: $E = - \hbar^2 / m\a^2 + (N-1)^2 \hbar\omega/2 + O(\gamma_t^{-1})$.

\begin{figure}
\includegraphics[width=\columnwidth, angle=0]{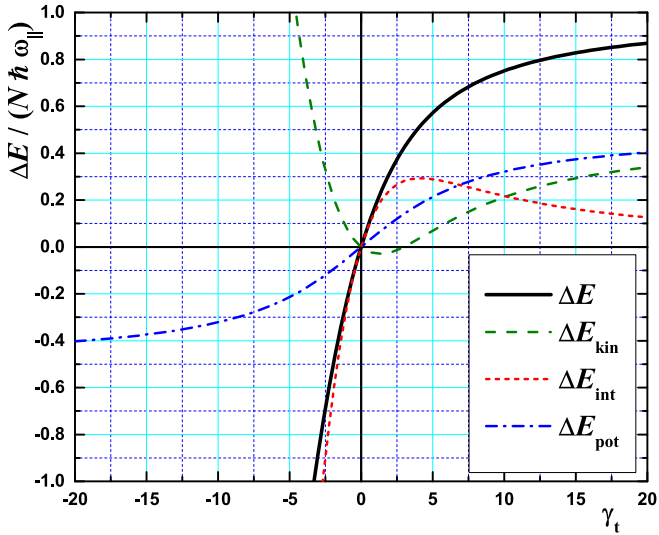}
\caption{Interaction shift to the energy of a trapped gas with an impurity.
Black solid line - total energy,  Eq.~(\ref{E:dim}),
blue short dashed line - potential energy of the harmonic confinement, Eq.~(\ref{Epot:dim}),
red dash-dotted line - interaction energy, Eq.~(\ref{Eint:dim}),
green dashed line - kinetic energy, Eq.~(\ref{Ekin:dim}).}
\label{Fig3}
\end{figure}

The interaction and kinetic energy can be obtained in a similar way as in the homogeneous system:
\begin{eqnarray}
\!\!\!\!\!\!\!\!\!\frac{E_{int}}{\hbar\omega_{\parallel}}
&\!\!=\!\!&
\frac{2 N\gamma_t}{\pi^2}\left(1-\frac{\gamma_t}{4}+\frac{\gamma_t}{2\pi}\arctg\frac{\gamma_t}{2\pi}\right)
\label{Eint:dim}\\
\!\!\!\!\!\!\!\!\!\frac{E_{kin}}{\hbar\omega_{\parallel}}
&\!\!=\!\!&
\frac{N^2\!+\!1}{4}
\!+\!\frac{N\gamma_t}{\pi^2}\!\left(\!
  \frac{\gamma_t}{4}\!-\!1\!+\!\left(\!\frac{\pi}{\gamma_t}
  \!-\!\frac{\gamma_t}{2\pi}\!\!\right)\!\arctg\!\frac{\gamma_t}{2\pi}\!
\right)
\label{Ekin:dim}
\end{eqnarray}
In Fig.~\ref{Fig3} we compare the different contributions to the energy shift in a trapped system.
The interaction energy vanishes for $\gamma_t =0$ or $\gamma_t \to +\infty$, as the system is equivalent to that of ideal fermions,
while for $\gamma_t \to - \infty$ it becomes essentially that of a dimer atom-impurity bound state. The potential energy contribution to the shift becomes substantial  when approaching  $\gamma_t \to \pm \infty$.

\begin{figure}
\includegraphics[width=\columnwidth, angle=0]{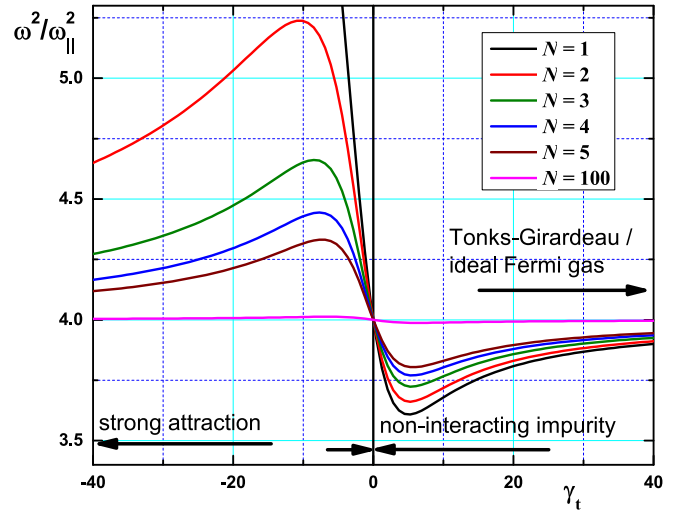}
\caption{Square of the frequency of the breathing mode as a function of $\gamma_t$ as given by Eq.~(\ref{omega2}).}
\label{Fig4}
\end{figure}

Another experimentally important quantity is the frequency of the breathing mode $\omega$.
It is related to the response of the system size on changing the frequency of the confinement $\omega^2 = -2 \langle x^2\rangle/(d\langle x^2\rangle/d\omega_{||}^2)$\cite{stringari}:
\begin{eqnarray}
\frac{\omega^2}{\omega^2_{||}}
=
\left(
\frac{1}{4}
+
\frac{N}{
  \left(\gamma_t+\frac{4\pi^2}{\gamma_t}\right)
  \left(N^2+1+\frac{4N}{\pi}\arcctg\frac{2\pi}{\gamma_t}\right)
}
\right)^{-1}
\label{omega2}
\end{eqnarray}
As can be seen from Fig.~\ref{Fig4}, $\omega$ changes significantly, only for small number of atoms. In the limit of infinite repulsion the impurity behaves as an additional fermion, leading to the ideal Fermi gas result $\omega^2/\omega^2_{||}=4$. Making the repulsion finite, the interaction ``softens'' and the frequency goes down.
In three limiting cases ($\gamma_t\to 0; \pm\infty$) the result is the same as for ideal fermions making the overall dependence non-monotonic. For weak interactions, $\gamma_t\to 0$, Gaussian ansatz result~(\ref{E:weak interaction:expansion}) is applicable leading to a linear dependence with $\gamma_t$: $\omega^2/\omega^2_{||}=4-4\gamma_t/(\pi^2N)$.
Also the expansion for strong interactions $\gamma_t\to\pm\infty$ $\omega^2/\omega^2_{||}=4-16N/[(N\pm 1)^2\gamma_t-4N]$ decays as $1/N$ for large system sizes.

In conclusion, we provided a theoretical treatment for an ideal Fermi gas in a presence of a single impurity interacting with arbitrarily strong $\delta$-potential.
In a homogeneous system the total energy was obtained by McGuire\cite{mcguire}. Here we analyze potential and kinetic energy, as well as experimentally relevant impurity-fermion local correlation function\cite{kinoshitag2}. In a trapped system we generalize McGuire's formula and demonstrate in a direct comparison with DMC result that the obtained expression is accurate. This expression in used to calculate kinetic, interaction and potential energies.
Finally, the frequency of the breathing mode is reported for different number of fermions.


GEA acknowledges fellowship by MEC (Spain) through the Ramon y Cajal fellowship program and financial support by (Spain) Grant No.
Part of the numerical calculations was carried out at Barcelona Supercomputing Center.
IB acknowledges financial support from the EC funded FET project QIBEC, and is very thankful to the experimental group of Selim Jochim for invaluable discussions and bringing into attention Ref.~\cite{mcguire}.

\end{document}